\documentclass[preprint,12pt,times]{elsarticle}

\usepackage{lineno}
\usepackage{graphicx}
\usepackage{epstopdf}
\usepackage{amssymb}
\usepackage{amsfonts,amsmath,bm,bbm}
\usepackage{mathtools}
\usepackage{enumerate}
\usepackage{indentfirst}
\usepackage[font=footnotesize]{caption}
\usepackage{floatrow}
\usepackage{enumitem}
\usepackage[table,xcdraw,dvipsnames]{xcolor}
\usepackage{url}
\usepackage{multirow}
\usepackage{cite}
\usepackage{floatrow}
\usepackage{booktabs} 
\usepackage{placeins}
\usepackage{multicol}
\usepackage{enumitem}
\usepackage{hyperref}
\usepackage{subcaption}
\usepackage{tabularray}
\usepackage{tabularx}
\usepackage[margin=2.5cm]{geometry}

\journal{Energy Conversion and Management}

\begin{document}

\begin{frontmatter}

\title{Statistical and economic evaluation of forecasts in electricity markets: beyond RMSE and MAE}

\author[ORBI]{Katarzyna Maciejowska}
\ead{katarzyna.maciejowska@pwr.edu.pl}
\author[ONAS]{Arkadiusz Lipiecki}
\ead{arkadiusz.lipiecki@pwr.edu.pl}
\author[ORBI]{Bartosz Uniejewski\corref{cor1}}
\ead{bartosz.uniejewski@pwr.edu.pl}
\cortext[cor1]{Corresponding author: Bartosz Uniejewski}

\address[ORBI]{Department of Operations Research and Business Intelligence}
\address[ONAS]{Department of Computational Social Science}
\address{Wrocław University of Science and Technology, Poland}

\begin{abstract}
Electricity price forecasts are typically evaluated using accuracy measures such as RMSE and MAE, although these metrics often fail to reflect their economic value in operational decisions. This paper investigates which statistical properties of electricity price forecasts are most relevant for economic performance, using battery energy storage system (BESS) arbitrage as an application.
We assess prediction quality along four dimensions: forecast accuracy, intraday error dispersion, association between predicted and realized prices, and the ability to identify daily price extrema. We construct a comprehensive pool of 192 hourly day-ahead electricity price forecasts and use it to evaluate the relationship between proposed quality measures and profits generated for two representative BESS configurations.
The results show that traditional accuracy metrics are only weakly correlated with BESS income. At the same time, dispersion- and association-based measures better capture a forecast’s economic value by reflecting its ability to reproduce daily price patterns. These findings demonstrate that incorporating complementary evaluation criteria may improve forecast selection and enhance the economic performance of BESS.
\end{abstract}


\end{frontmatter}

\FloatBarrier
\begin{table*}
\caption*{\large{Glossary}}

\begin{tabular}{|c|l|}
  \hline
  ARX & AutoRegressive model with eXogenous variables \\
  \hline 
  BESS & Battery Energy Storage Systems \\
  \hline
  Corr-f & Average correlation between the observations and forecasts \\
  \hline
  Cov-e & Logarithm of a determinant of the covariance matrix of forecast errors \\
  \hline
  DA & Day-Ahead electricity market \\
  \hline
  DFL & Decision-Focused Learning \\
  \hline
  LEAR & LASSO-Estimated AutoRegressive model \\
  \hline
  MAE & Mean Absolute Error \\
  \hline
  MHD & Min-Max Hour Deviation \\
  \hline
  MPD & Min-Max Price Deviation \\
  \hline
  NARX & Nonlinear AutoRegressive model with eXogenous variables \\
  \hline 
  RES & Renewable Energy Sources \\
  \hline
  RMSE & Root Mean Square Error \\
  \hline
  VST & Variance Stabilizing Transformations \\
  \hline
  $B$ & Length of a charging block (hour) \\
  \hline
  $C$ & Operating costs (EUR/MWh) \\
  \hline 
  $E$ & Battery energy capacity (MWh) \\
  \hline
  $Pow$ & Battery power rating (MW) \\
  \hline
  $\eta^{ch}$ & Charging efficiency (\%) \\
  \hline
  $\eta^{dis}$ & Discharging efficiency (\%) \\
\hline
  
\end{tabular}
\end{table*}

\FloatBarrier

\section{Introduction}
Forecasting electricity prices is vital for the efficient functioning of modern energy markets. For producers, accurate forecasts support generation planning and trading strategies, while large consumers use them to optimize energy use and reduce costs. Methods range from simple regression and time-series models to advanced approaches such as decision trees and neural networks, often enhanced through postprocessing techniques, such as forecast combination, to improve accuracy~\citep{lag:mar:des:wer:21, jed:etal:2022, tsc:etal:2022}.

Despite this diversity, it remains unclear which forecasts provide the greatest value for decision support. Forecast selection is commonly based on statistical accuracy measures such as the root mean squared error (RMSE) or mean absolute error (MAE) {, which directly quantify the deviation between forecasts and realized observations. Although perfect foresight -- corresponding to zero RMSE and MAE -- would lead to optimal decisions, numerous studies show that in empirical applications better statistical accuracy does not necessarily translate into greater economic value (see \citealp{mur:1993, zar:can:bha:10,lin:zhu:wid:24}, among others,).
One key reason is the symmetry of conventional accuracy metrics, which fails to account for the asymmetric costs associated with over- and under-prediction \citep{keb:ara:rah:2011,li:chi:2018,zha:wan:hug:2022,ser:wer:2024}. Moreover, such measures typically ignore the joint impact of forecast errors on downstream decision problems. While the literature increasingly acknowledges the mismatch between minimizing RMSE or MAE and maximizing economic performance, there remains a lack of systematic and comprehensive research addressing this issue in the context of energy markets.}

In order to relate forecasting and optimization, \emph{decision-focused learning} (DFL) techniques have been proposed in the literature. The DFL approach directly incorporates cost or profit functions into the estimation stage, thereby aligning the forecasting objective with downstream decision-making \citep{Zha:23,mand:etal:24,car:kar:19}. Recent applications of DFL in energy markets report promising results and often outperform the traditional two-stage predict--then--optimize framework \citep{par:etal:25,alk:etal:26}. However, the benefits of DFL are application-dependent and, as shown by \citep{mand:etal:24}, do not necessarily translate into higher economic value, while typically requiring substantially greater computational effort.

Although value-oriented methods can be advantageous, several barriers limit their adoption in energy companies. Forecast are typically provided by third-party vendors, who sell the same predictions to various firms. Even within a single company, there are used across multiple departments, such as generation, trading, and risk management. The differing objectives and cost structures faced by these end users make the specification of a single loss function within a decision-focused learning framework impossible. Finally, predictions are frequently evaluated by human experts who select the final forecast or make operational decisions; in such settings, forecast targets must be easy to interpret (e.g., mean, median, or quantile). By contrast, the minimizer of a cost-oriented loss function may lack a clear or intuitive interpretation \citep{car:kar:19}.

As a result, rather than modifying the calibration process -- as in decision-focused learning -- firms may instead prioritize adjusting the forecast selection method. Accordingly we focus on identifying statistical properties of electricity price forecasts that are particularly useful for the management of battery energy storage systems (BESS). 
Specifically, we evaluate prediction quality from four perspectives: (i) forecast accuracy, measuring the distance between forecasts and realized values, as captured by RMSE and MAE; (ii) error dispersion, reflecting the degree of diversification of forecast errors within a day; (iii) the association between predicted and realized prices; and (iv) the ability of forecasts to correctly identify the hours corresponding to the daily minimum and maximum prices.

We select battery energy storage systems as an illustrative application for assessing the value of electricity price forecasts, because of their growing importance in modern power systems. BESS enables the temporal shifting of electricity by storing surplus energy during periods of high renewable generation and dispatching it during periods of elevated system demand. In doing so, they enhance grid reliability and improve overall energy management. The economic viability of BESS deployment is primarily driven by four revenue streams: price arbitrage, load shifting, frequency regulation, and voltage support \citep{yam:etal:22,sch:sta:23,sha:26}. In this study, we focus on arbitrage that explores temporal price differentials, as this revenue source depends directly on the quality of predictions. 

To bridge forecast quality measures and economic value, we construct a comprehensive pool of 192 hourly day-ahead price forecasts. These include predictions from three common model types: autoregressive with exogenous variables (ARX), its nonlinear variant based on artificial neural networks (NARX), and the regularized LASSO-estimated autoregressive model (LEAR). The pool is further diversified through variations in model specifications, variance-stabilizing transformations, estimator types, and calibration window sizes. Each of the forecasts is evaluated with seven forecast quality measures and the profits generated by two exemplary BESS.

The results provide strong evidence that traditional accuracy measures are only weakly correlated with BESS profits, confirming earlier findings of \citep{zar:can:bha:10}. In contrast, dispersion- and association-based measures more effectively capture the extent to which forecasts reproduce the intraday price profile and, consequently, their ability to identify profitable charging and discharging periods. These findings indicate that restricting forecast evaluation to accuracy metrics alone limits the selection of economically valuable models, whereas incorporating complementary evaluation criteria may improve forecast selection and enhances BESS profitability.
 
The major contributions of this study are: 
\begin{enumerate}
\itemsep = 0pt
       \item A comprehensive panel of 192 forecasts based on statistical and machine learning models is constructed, incorporating different data transformations, model specifications, and calibration window selections. The results indicate that the BESS profits depend heavily on the selected forecast.
    \item We provide evidence that classical forecast accuracy measures, such as the RMSE or the MAE, are only weakly correlated with BESS profits. 

    \item We propose and examine two measures describing dispersion of errors and association of forecasts, which reflect statistical properties of the electricity price forecasts that are particularly relevant for the BESS decision-making process. The results indicate a strong correlation between these measures and battery profits.
\end{enumerate}


The paper is structured as follows. In Section \ref{sec:data}, we present and describe the dataset; in Section \ref{sec:forecast}, we explain the forecasting models applied in this study. Then, in Section \ref{sec:measures}, we lay out the methodology for evaluating the forecasts, both in terms of economic values and statistical accuracy. Next, in Section \ref{sec:result}, we present the empirical results of our forecasting and trading exercises and discuss its limitations in Section \ref{sec:Discussion}. Finally, Section \ref{sec:conclusion} summarizes the main results and provides recommendations for forecasters aiming to capture profits in day-ahead electricity markets.

\section{Data}
\label{sec:data}

\begin{figure}[tb!]
 \centering
 \includegraphics[width = \textwidth]{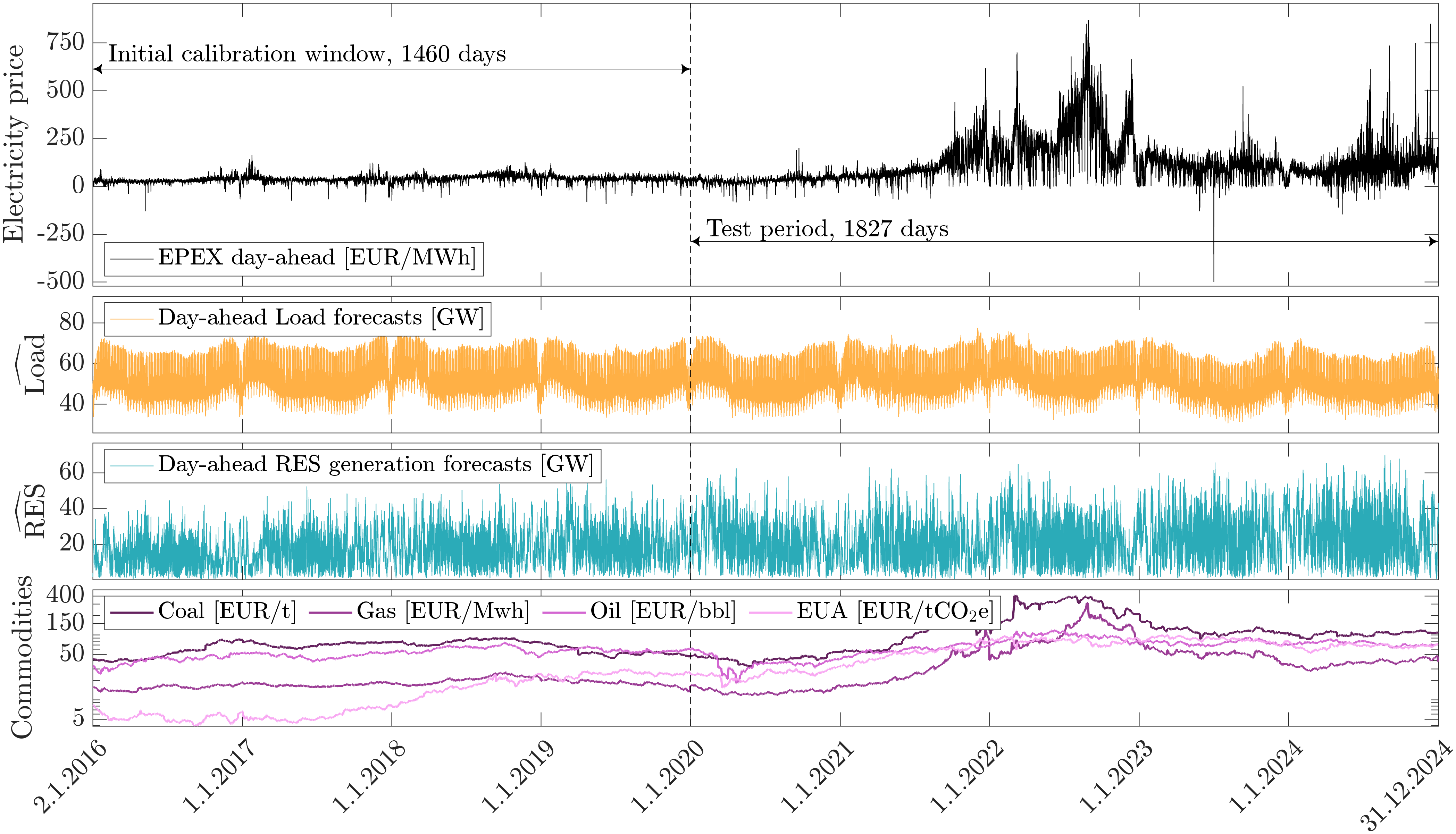}
 \caption{EPEX SPOT hourly day-ahead electricity prices (top), hourly day-ahead system load forecasts (upper middle), renewable generation from solar and wind sources (lower middle), and commodity prices are shown for the period from 2.1.2016 to 31.12.2024. The vertical dashed line marks the end of the 1460-day calibration window for the forecasting models and the beginning of the 2199-day out-of-sample test period.}
 \label{fig:data}
\end{figure}

To examine the link between statistical forecast quality and economic value, we use the German electricity market -- one of Europe’s largest and most active -- as a case study. Specifically, we assess how day-ahead electricity price forecasts influence the profitability of a battery energy storage system (BESS) engaged in daily arbitrage.

Our dataset combines key drivers of electricity prices and market behavior. The core series is the hourly day-ahead price for the Germany–Luxembourg bidding zone (Austria included until October 2018), sourced from the ENTSO-E transparency platform. To capture supply and demand fundamentals, we add ENTSO-E day-ahead forecasts of system load, solar generation, and aggregated wind generation (onshore and offshore), with solar and wind combined into a single renewable generation series. Broader market influences are represented by commodity futures prices for coal (API2), natural gas (TTF), crude oil (Brent), and EU emission allowances (EUA), obtained from Investing.com. All series cover 2.1.2016–31.12.2024, with a five-year out-of-sample period (see Figure~\ref{fig:data}). The out-of-sample period starts in 2020 to explicitly include major structural changes in electricity prices related to the COVID-19 pandemic and the subsequent energy crisis.

Data were preprocessed for temporal and structural consistency. Variables originally reported at 15-minute resolution (e.g., load and renewables) were aggregated to hourly frequency. Daylight saving time transitions were also adjusted: missing values during the spring shift to CEST were imputed by averaging adjacent hours, while duplicated hours during the autumn return to CET were replaced with their mean.

\section{Forecasting electricity prices}
\label{sec:forecast}

\subsection{ARX model}

In this research, we use the Autoregressive Model with Exogenous variables (ARX) popular in the EPF literature \citep{zie:wer:18,lag:mar:des:wer:21, mac:uni:wer:23}. Since in the DA market, all prices are set at the same time via a simultaneous auction, the data does not have a typical time-series structure and therefore each hour is modeled separately. Let us denote by $Y_{t,h}$ the dependent variable on day $t$ and at hour $h$, which differs across model specifications. We adopt the following structure:
\begin{equation}
\label{eq:ARX}
    Y_{t,h} = D_t\alpha_h + \sum_{p\in \{1,2, 7\}}Y_{t-p,h}\rho_{p,h} + X^{(1)}_{t}\beta_{1,h} +X^{(2)}_{t,h}\beta_{2,h}+ \varepsilon_{t,h},
\end{equation}
where $D_t$ is a $(1\times 3)$ vector of deterministic variables that  includes a constant and dummies for Weekends and Mondays. The variable $Y_{t-p,h}$ is a lagged endogenous variable at hour $h$ from $p$-days ago, and $X^{(1)}_{t}$, $X^{(2)}_{t,h}$ are vectors of exogenous variables. $X^{(1)}_{t}$ is a $(1\times 9)$ vector that summarizes information common to all hourly contracts:
\begin{itemize}
    \item information on previous day prices: $\min_h(P_{t-1,h})$, $\max_h(P_{t-1,h})$, $\bar{P}_{t-1}$
    \item information on the average forecasted level of RES and Load: $\bar{RES}_t$ and $\bar{L}_t$
    \item fuel and $CO_2$ allowance prices from day $t-2$: gas ($G_{t-2}$), oil ($Oil_{t-2}$), coal ($C_{t-2}$) and EUA ($EUA_{t-2}$).
\end{itemize}
The vector $X^{(2)}_{t,h}$ consists of information characteristic for an hour $h$. It includes information about the TSO predictions of Load and RES generation from the current and previous day: $L_{t,h}$, $RES_{t,h}$, $L_{t-1,h}$, $RES_{t-1,h}$. The selection of exogenous variables differs slightly from popular ARX model specification (\citep{mac:uni:wer:23}). The set is extended to accommodate recent results of  \citep{mar:etal:23}, which show the importance of past information on generation structure in the price forecasting process.

\subsubsection{Direct model of hourly prices}

ARX model is fitted to the hourly electricity prices: 
\begin{equation*}
    P_{t,h} = D_t\alpha_h + \sum_{p\in \{1,2, 7\}}P_{t-p,h}\rho_{p,h} + X^{(1)}_{t}\beta_{1,h} +X^{(2)}_{t,h}\beta_{2,h}+ \varepsilon_{t,h}.
\end{equation*}
The model is characterized by 16 parameters: $\theta_h = [\alpha_h', \rho_{1,h}, \rho_{2,h}, \rho_{3,h}, \beta'_{1,h},\beta'_{2,h}]'$ and is estimated with the Ordinary Least Squares (OLS) method.

\subsubsection{Model of deviation from daily mean}
In this paper, we consider also an alternative ARX-type model that describes separately the average daily price and the deviation of hourly prices from the daily mean. 
The approach is divided into two steps: forecasting the daily average ($\bar{P}_t$) and predicting the deviation from the mean
\begin{equation}\label{eq:deviation:mean}
\tilde{P}_{t,h} = P_{t,h} - \bar{P}_t.
\end{equation}

The model for the daily average takes the following form:
\begin{equation*}
    \bar{P}_t = D_t\alpha + \sum_{p\in \{1,2, 7\}}\bar{P}_{t-p}\rho_{p} + X^{(1)}_{t}\beta_{1} +\bar{X}^{(2)}_{t}\beta_{2}+ \varepsilon_{t},
\end{equation*}
where the vector $\bar{X}^{(2)}_{t}$ includes the average daily values of the previous day's generation structure:  $\bar{L}_{t-1}$, $\bar{RES}_{t-1}$.

The model for deviation from the mean is designed analogously, but with hour resolution
\begin{equation*}
    \tilde{P}_{t,h} = D_t\alpha_h + \sum_{p\in \{1,2,7\}}\tilde{P}_{t-p,h}\rho_{p,h} + X^{(1)}_{t,h}\beta_{1,h} +X_{t,h}^{(2)}\beta_{2,h}+ \varepsilon_{t,h},
\end{equation*}
where the vector $X^{(2)}_{t,h}$ is the same as in the direct model of hourly prices and includes information on the generation structure from current and previous day: $ L_{t,h}$, ${RES}_{t,h}$, ${L}_{t-1,h}$, ${RES}_{t-1,h}$. 
The final prediction is computed as the sum of forecasts of both components.

\subsection{NARX}

The relationship between the lagged prices, exogenous variables, and the future price does not need to be linear. To forecast the prices without an explicit assumption about the transformation between the regressors and the target, we resort to artificial neural networks. To ensure that the results of our study are relatable to popular forecasting practices, we employ a shallow network consisting of a single hidden layer of 5 neurons with hyperbolic tangent activation functions, an architecture used in many previous EPF studies \citep{mar:uni:wer:19:narx, hub:mar:wer:19, mar:uni:wer:20}. The model is trained using a Levenberg-Marquardt algorithm~\citep{hag:men:94} with a random 10\% of the training window held out for early stopping. To reduce the uncertainty related to parameter estimation, the results presented in the paper correspond to the committee machine approach~\citep{mar:uni:wer:19:narx}, in which a final prediction is an ensemble average of predictions obtained from ten independently trained neural networks. The regressors, calibration window lengths, and applied variance stabilizing transformations are the same as in the case of ARX models.

\subsection{LEAR model}
The next model considered is the LASSO-Estimated AutoRegressive (LEAR) model. It employs the Least Absolute Shrinkage and Selection Operator (LASSO) introduced by Tibshirani \citep{tib:96} to automatically select the most relevant predictors for forecasting $Y_{d,h}$. While numerous regularization techniques have been proposed in the literature, Uniejewski \citep{uni:24:ORD} identified LASSO as a particularly parsimonious, robust, and high-performing method in a comprehensive evaluation of electricity price forecasting (EPF) models.

The LEAR model (both direct and for modeling the deviation from the mean) is designed to capture extensive cross-hour dependencies, enhancing its ability to represent the temporal structure of electricity prices. In contrast to the ARX model defined in Eq.~\eqref{eq:ARX}, which includes only three autoregressive terms, the LEAR specification expands this component to include past prices from all 24 hours of the previous day, two days prior, and one week prior—replacing three regressors with a total of 72. Including all 24 hourly prices from the previous days reflects the cross-hour dependencies as well as daily and weekly seasonal patterns in electricity prices.

A similar expansion is applied to the exogenous input vector $X^{(2)}_{t,h}$. Instead of four regressors (two each for load and RES day-ahead forecasts), the LEAR model incorporates 96 predictors, capturing values for every hour of both the forecasted day and the previous day. This comprehensive treatment of temporal dependencies makes the LEAR model well-suited for capturing the complex dynamics of electricity price formation. Lastly, the model is expanded to include all seven dummy variables, one for each day of the week.

\subsection{Model specifications and forecast averaging}

To develop our forecasting methodology, we begin with three core model structures: ARX, NARX, and LEAR. Each model is estimated across multiple variants to account for different modeling assumptions and data transformations. These variants differ in the choice of the dependent variable, the level of parameter estimation (either pooled across all hours or separate for each hour, already describe above), the size of the calibration window, and the application of a variance stabilizing transformation. The specific choices and their implementation are described in detail in the following sections.

\subsubsection{Size of a calibration window}

We consider seven estimation window sizes: three long windows (one, two, and four years), a medium window of 182 days, and three short windows of six, twelve, and sixteen weeks. As shown by \citep{mar:ser:wer:18}, no single calibration window is universally optimal in electricity markets. Long windows reduce estimator variance and, under stationarity, improve precision. However, structural changes (e.g., rising RES shares) and exogenous shocks (COVID-19, the war in Ukraine) often break stationarity, while price–fundamental relationships are frequently nonlinear. In such cases, short windows can be advantageous. Prior studies \citep{hub:mar:wer:19, ser:uni:wer:19} further suggest that combining forecasts across different window lengths yields the best accuracy.

\subsubsection{VST data transformation}
Electricity price spikes are typically driven by unpredictable weather conditions, power outages, or transmission failures \citep{gia:gro:12}. These extreme events can significantly distort electricity price forecasts, as outliers tend to pull model coefficients toward values that fit the spikes, often at the expense of forecast accuracy during normal periods. Variance stabilizing transformations (VSTs) aim to reduce the overall variability in the data \citep{cia:mun:zar:22}, and less variable—or less spiky—input data generally enable forecasting models to produce more accurate predictions \citep{jan:tru:wer:wol:13}.

Following the approach of \citep{uni:wer:zie:18}, the electricity price and load time series are first standardized by subtracting the sample median $a$ and dividing by the sample median absolute deviation $b$. The area hyperbolic sine (asinh) transformation is then applied to the standardized data. After forecasting in the transformed space, the inverse transformation and standardization are applied to obtain the final electricity price forecasts.

\subsubsection{Heterogeneous vs. pooled estimator}
{We consider two types of estimators.} Firstly, each hour is modeled and predicted individually, so estimators of $\theta_h$ and $\tilde{\theta}_h$ change throughout the day. In the remaining part of the article, {we call it} a \textit{heterogeneous} estimator as it is able to accommodate differences in price behavior across hours. Next, a \textit{pooled} estimator is used that assumes that the response of prices to explanatory variables is constant during a day, so $\theta = \theta_1=\ldots = \theta_{24}$. The vector of parameters is estimated using market information from all 24 hours at once. It is particularly suitable for short estimation windows as it provides more data to the calibration algorithm.  

Note that the LEAR model, which incorporates observations for all 24 hours of the day, uses the same set of input variables regardless of the hour being forecasted. As a result, the pooled estimation approach must be supplemented with additional variables that capture hour-specific dynamics. Compared to the heterogeneous estimation (which estimates separate models for each hour), the pooled model is augmented with seven additional regressors: electricity prices lagged by 24, 48, and 168 hours, and load and RES forecasts lagged by 0 and 24 hours.

\subsubsection{Forecast averaging}

Forecast averaging is a very powerful post-processing method that improves the accuracy of predictions \citep{arm:01, tim:06, ati:2020}. In the  energy price forecasting literature, many different approaches have been discussed. In \citep{mac:now:wer:16,lip:etal:2024} authors combine predictions obtained with different models,  in \citep{ser:uni:wer:19} and \citep{hub:mar:wer:19}  a single model is used and fitted to calibration windows of different sizes. Finally, \citep{mar:etal:23}  computes the average across outcomes of different realizations of neural networks. Here, we adopt the second approach and include in the pool of predictions averages of forecasts over different sizes of estimation windows.

\subsubsection{Summary}

For each hour in the evaluation period, we prepare a pool of predictions that come from different models, model specifications, and forecast averaging. There are three models (ARX, NARX, LEAR) and eight model specifications that differ in terms of the dependent variable (models of hourly prices, $P_{t,h}$, or deviation from the daily mean, $\tilde{P}_{t,h}$), VST transformation, and type of estimator (heterogeneous, pooled). Finally, parameters of each model are estimated with data belonging to windows of different lengths: 56, 84, 112, 182, 365, 730, and 1460 days. {A comprehensive description of models included in the pool is presented in Table \ref{tab:pool:summary}.}

We consider four different pools of predictions. The first one is based on the results of ARX models and consists of 56 individual predictions and 8 forecast averages. Hence, it is built up from 64 different predictions. Analogously, pools based on NARX and LEAR models are constructed. 
Finally, a large pool that consists of predictions obtained with all analyzed models, model specifications and corresponding forecast averages is analyzed. It consists of 192 forecasts and captures the diversity stemming from both, the type of a model and its specification.

\begin{table}[]
{
    \caption{Description of models included in the pool}
    \centering
    \begin{tabularx}{\textwidth}{X *{6}{|>{\centering\arraybackslash}m{16mm}}}
         Specification & \multicolumn{6}{>{\centering\arraybackslash}m{96mm+10\tabcolsep}}{Values} \\
         \hline
         \hline
         Model & \multicolumn{2}{>{\centering\arraybackslash}m{32mm+2\tabcolsep}|}{ARX} & \multicolumn{2}{>{\centering\arraybackslash}m{32mm+2\tabcolsep}|}{NARX} & \multicolumn{2}{>{\centering\arraybackslash}m{32mm+2\tabcolsep}}{LEAR} \\
         \hline
         VST & \multicolumn{3}{>{\centering\arraybackslash}m{48mm+4\tabcolsep}|}{None} & \multicolumn{3}{>{\centering\arraybackslash}m{48mm+4\tabcolsep}}{Asinh} \\
         \hline
         Estimator type & \multicolumn{3}{>{\centering\arraybackslash}m{48mm+4\tabcolsep}|}{Heterogeneous} & \multicolumn{3}{>{\centering\arraybackslash}m{48mm+4\tabcolsep}}{Pooled} \\
         \hline
         Dependent variable & \multicolumn{3}{>{\centering\arraybackslash}m{48mm+4\tabcolsep}|}{Hourly prices} & \multicolumn{3}{>{\centering\arraybackslash}m{48mm+4\tabcolsep}}{Daily mean \& hourly deviations} \\
         \hline
         Window size & \multicolumn{3}{>{\centering\arraybackslash}m{48mm+4\tabcolsep}|}{56, 84, 112, 182, 365, 730, 1460} & \multicolumn{3}{>{\centering\arraybackslash}m{48mm+4\tabcolsep}}{Average over all windows} \\
         \hline
    \end{tabularx}
    \label{tab:pool:summary}
}
\end{table}

\section{Forecast evaluation}
\label{sec:measures}

According to \citep{mur:1993}, there are three distinct types of forecast goodness: consistency (Correspondence between forecasters' judgments and forecasts), quality (Correspondence between the forecasts and observations), and value (benefits from using the forecasts). The goal of this research is to evaluate the link between the quality and the value of electricity price point forecasts. The study is based on the analysis of the economic performance of a Battery Storage System (BESS). 

\subsection{Economic value of forecasts }
To assess the economic value of predictions, we analyze the performance of a BESS system that places orders in the DA market. 
Similar to \citep{lin:zhu:wid:24}, we assume that the BESS system is based on the grid-scale Lithium-ion batteries. The specification of an exemplary BESS is summarized by Table \ref{tab:BESS} that shows parameter values for two hypothetical systems: BESS-a and BESS-b. Both batteries are assumed to have the same energy capacity of 3 MWh. They differ in terms of the power ratings that are 3 MW and 1 MW, for BESS-a and BESS-b respectively. It influences the charging and discharging speed of BESS. The first system needs one hour, whereas the second one requires three hours to fully charge or discharge. {The operating costs and battery efficiency are taken from \citep{lin:zhu:wid:24} and represent the levelled cost per unit of energy throughput. This takes into account investment and O\&M costs, as well as implicitly incorporating battery degradation through the assumed number of charge–discharge cycles over the lifetime of the battery.}

\begin{table}[h]
    \caption{Specification of BESS}
    \label{tab:BESS}
    \centering
    \SetTblrInner{rowsep=0.5pt}
 \begin{tblr}{
  colspec={l|c|cc|c},
  cell{1}{1} = {c=2}{l},
}
     Parameter  & & BESS-a & BESS-b & Unit \\
    \hline
    \hline
    Energy capacity & $E$ & 3 & 3 & MWh\\
    Power rating& $Pow$ & 3 & 1 & MW\\
    Charging block & $B$ & 1 & 3 & hour\\
    \hline
    Charging efficiency&  $\eta^{ch}$ & 98  & 98 & \%\\
    Discharging efficiency &  $\eta^{dis}$ & 97  & 97 & \%\\
    \hline
    Operating costs & $C$ & 11.63 & 11.63 & EUR/MWh\\
    \hline
\end{tblr}  
\end{table}

It is assumed that BESS earns income from the arbitrage in DA market. BESS aims at buying energy in periods of low prices and selling at hours of high prices. To plan the operation, BESS needs to choose one day ahead when to trade the electricity. {Here, we} assumed that the charging or discharging is processed in consecutive hours (blocks of 1-3 hours). The hours are selected using the forecast of DA prices. Let us denote by $h^{ch}$ and $h^{dis}$ the selected first hour of charging and discharging blocks, respectively. Then the profit earned on the day $t$ can be calculated as
\begin{equation}
    \pi^E_t = \eta^{dis}\sum_{i=0}^{B-1}{Pow}^{dis}_{i+h^{dis}}P_{t,i+h^{dis}}  - \frac{1}{\eta^{ch}}\sum_{i=0}^{B-1} {Pow}^{ch}_{i+h^{ch}}P_{t,i+h^{ch}}-2CE,
\end{equation}
where ${Pow}^{ch}_{h}$ and ${Pow}^{dis}_{h}$ is power charged and discharged within a particular hour, and $B$ is the length of a charging block. Notice that due to the BESS specification, following constraints need to be satisfied ${Pow}^{ch}_{h}\leq {Pow}$, ${Pow}^{dis}_{h}\leq {Pow}$ and 
$$\sum_{i=0}^{B-1} {Pow}^{ch}_{h+i}\leq E,\quad\sum_{i=0}^{B-1} {Pow}^{dis}_{h+i}\leq E.$$
Finally, {we assumed} that the battery runs a full cycle within a day; hence, no extra energy is carried over night to be used the next day. Additionally, since the level of profits depends on the energy capacity of the battery, in the following part of the article, we report a profit per 1 MWh of trade, which is calculated as $\pi_t = \pi^E_t/E$.

\subsection{Quality of forecasts}

In the literature, the quality of electricity price forecasts is typically described by metrics that focus on forecast accuracy such as Root Mean Squared Errors (RMSE) or Mean Absolute Errors (MAE). They are based on forecast errors of predictions from individual hours, $e_{t,h} =P_{t,h} - \hat{P}_{t,h}$. In order to obtain a single measure that describes the performance across 24 hours, results are usually combined into daily quantities. Let us denote by $e_t = [e_{t,1}, \dots,e_{t,24}]$ a $(1\times 24)$ vector of forecast errors on day $t$. Then
\begin{equation*}
    \text{RMSE} = \sqrt{\frac{1}{24T}\sum_{t=1}^T||e_t||_2^2}\;,\quad\text{MAE} = \frac{1}{24T}\sum_{t=1}^T||e_{t}||_1
\end{equation*}
These two traditional accuracy metrics indicate the average discrepancy between predicted and actual values. They are commonly used because they align with widely adopted loss functions in model calibration. For instance, RMSE corresponds to the loss function associated with the Least Squares estimation method, while MAE aligns with the loss function in quantile regression (specifically for estimating the median). 
 
\begin{figure}

\centering

\includegraphics[width=\textwidth]{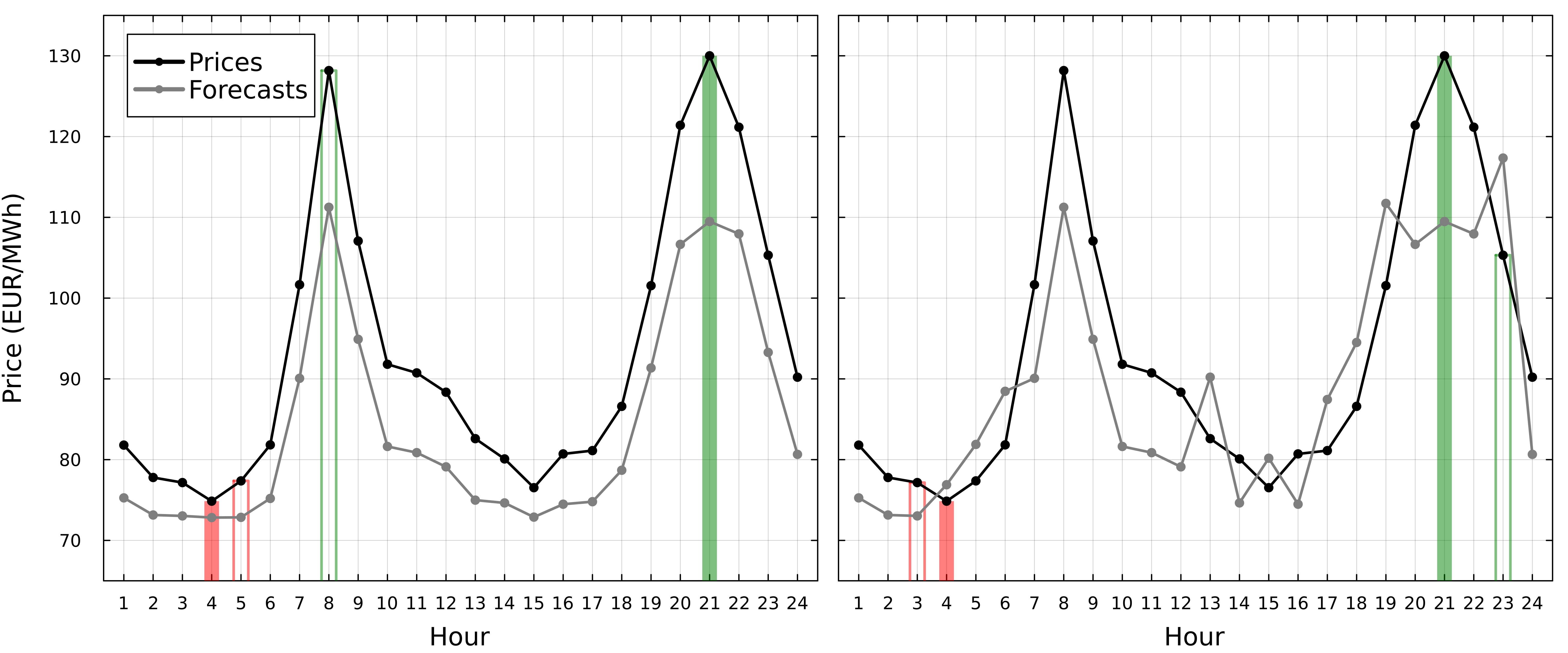}

\caption{Examples of DA price forecasts. \textbf{Setup 1} (left panel): forecasts with low dispersion and high association, leading to $\pi^{(1)}$ = 24.68 EUR profit. \textbf{Setup 2} (right panel): forecasts with high dispersion and low association, leading to $\pi^{(2)}$ = 0.16 EUR profit. Both predictions are characterized by the same RMSE and MAE values.}
\label{fig:mesures:example}
\end{figure}

Despite their popularity, accuracy metrics capture only a narrow dimension of forecast performance. Fig.~\ref{fig:mesures:example} illustrates this with two day-ahead forecasts that achieve identical values ($\text{RMSE}=10$, $\text{MAE}=9$) but lead to very different profits: $\pi^{(1)}=24.68$ in the first panel versus $\pi^{(2)}=0.16$ in the second. From a battery owner’s perspective, their economic value is therefore fundamentally different.

The forecasts in Fig.~\ref{fig:mesures:example} differ markedly. While their errors have similar magnitudes, their variability is not. The left panel tracks the daily pattern of day-ahead (DA) prices more closely than the right, making it more useful for identifying optimal charging and discharging hours. To capture such differences, we introduce three additional classes of metrics: those assessing forecast error variability, the association between forecasts and actual prices, and the ability to correctly identify the daily minimum and maximum price hours.

\subsubsection{Dispersion measures}
Dispersion measures show how diversified the forecast errors are within a day. If all errors share the same sign, as illustrated in Fig.~\ref{fig:mesures:example}, the predictions might not be precise, yet they are closely related to actual values. At an extreme, when errors are uniform, predictions can be simply viewed as shifted true observations.  Although not accurate, such forecasts may be highly beneficial in determining the optimal time for charging and discharging BESS.

In this study, we measure the dispersion using the variance-covariance matrix of forecast errors. It is calculated as
\begin{equation*}
    \text{Cov-e} = \log{det\hat{\Sigma}},
\end{equation*}
where 
\begin{equation*}
    \hat{\Sigma} = \frac{1}{T}\sum_{t=1}^Te'_{t}e_t.
\end{equation*}
Notice that if the predictions $\hat{P}_{t,h}$ are unbiased and forecast errors have zero expected value, then $\hat{\Sigma}$ is a consistent estimator of the errors' variance-covariance matrix. Moreover, Cov-e decreases as forecast errors decline and as the correlation between errors increases. Thus, it reflects not only the dispersion but also, to some extent, the overall forecast accuracy.

\subsubsection{Association measure}

According to \citep{mur:1993}, the association is the relationship between individual pairs of forecasts and observations. For instance, the daily price curve in left panel in Fig.~\ref{fig:mesures:example} more closely mirrors the actual price fluctuation pattern compared to the curve shown in right panel in Fig.~\ref{fig:mesures:example}. 
{Here, we measured it} as the average correlation between actual and predicted electricity prices. It is computed as follows
\begin{equation*}
    \text{Corr-f} = \frac{1}{T}\sum_{t=1}^T \rho(P_{t}, \hat{P}_t),
\end{equation*}
where $P_{t}$ and $\hat{P}_{t}$ are $(1\times 24)$ vectors of electricity prices and their forecasts on day $t$ and $\rho(.)$ is the Spearman correlation coefficient.

\subsubsection{Selection of an hour of the minimum and maximum price}

The final two measures capture forecast properties directly relevant for BESS operation: the correct identification of daily minimum and maximum price hours using point forecasts $\hat{P}_t$. Let $\hat{h}_t^{(min)}$ and $\hat{h}_t^{(max)}$ denote the forecasted hours of the lowest and highest prices on day $t$, and $h_t^{(min)}$, $h_t^{(max)}$ their actual counterparts from observed prices $P_t$.
The first measure, the \textit{Min-Max Hour Deviation} (MHD), computes the average absolute difference between forecasted and actual hours:
\begin{equation*}
\text{MHD} = \frac{1}{T}\sum_{t=1}^T |h_t^{(min)} - \hat{h}_t^{(min)}| + |h_t^{(max)} - \hat{h}_t^{(max)}|.
\end{equation*}
From the perspective of BESS, the incorrect selection of the moment of charging or discharging has a significant impact on income only when there are substantial differences of prices in actual and the predicted hours. Therefore, we propose a second measure, based on the difference between the real maximum/minimum of price and the maximum/minimum indicated by forecasts called \textit{Min-Max Price Deviation} (MPD):
\begin{equation*}
\text{MPD} = \frac{1}{T}\sum_{t=1}^T |P_{t,h_t^{(min)}}-P_{t,\hat{h}t^{(min)}}| + |P_{t,h_t^{(max)}}- P_{t,\hat{h}_t^{(max)}}|.
\label{eq:MPD}
\end{equation*}
Since $P_{t,h_t^{(max)}}\geq P_{t,h_t^{(min)}}$ and $P_{t,h_t^{(min)}}\leq P_{t,h_t^{(max)}}$, MPD can be interpreted as the average deviation in daily price spreads. It is closely related to profit loss from misidentifying charging and discharging hours, though not identical. MPD ignores storage inefficiencies and does not capture the broader dependence of profits on multiple hours for systems with a C-rating $E/Pow < 1$. Thus, while strongly correlated with income, MPD does not measure profit loss directly.

A related concept was introduced by \citep{ama:etal:2023}, who proposed d-RMSE, based on the difference between actual peak demand and demand at the predicted peak hour. Unlike d-RMSE, MPD uses absolute errors instead of squared errors, directly reflecting lost profits from misidentifying daily extrema, and jointly evaluates both peak and trough values.

 In Fig.~\ref{fig:mesures:example} red and green bars indicate the charging and discharging hours, respectively. To assess the accuracy of hour selection, full bars represent hours chosen using actual prices, while empty bars correspond to those selected based on forecasts. The height of the bars indicates the actual price at the selected hour. {The results show} that neither forecast perfectly identifies the optimal charging and discharging times. When examining the deviation from the optimal hours, the first set of predictions shows greater discrepancies: $\text{MHD}^{(1)} = 6.5$ whereas $\text{MHD}^{(2)} = 1.5$. Since the profit depends on the level of prices more than on the time of trade, we calculate also the second measure: $\text{MPD}^{(1)} = 0.91$ and $\text{MPD}^{(2)} = 13.50$ for the first and second panels, respectively. The findings highlight that the second approach aligns more closely with actual profits, making it a more appropriate choice for forecast evaluation.

\section{Empirical results}
\label{sec:result}

{
\subsection{Forecast accuracy}

To assess the rationale for using the proposed forecasting pool, we first compare forecast accuracy across different models and model specifications using standard point forecast metrics, namely RMSE and MAE. As the pool in this study comprises nearly two hundred individual predictions, we report in Table~\ref{tab:rmse_mae} only the results for the ensemble forecasts, computed as the average of predictions across all considered calibration windows. A more detailed view of the pool is provided in Fig.~\ref{fig:ScatterProfits}, which presents the results for all considered models and specifications.

Table~\ref{tab:rmse_mae} reports the values of both accuracy measures, calculated over the full test sample. For each model class -- ARX, NARX, and LEAR -- the most accurate specification is highlighted in bold. The results indicate that the nonlinear NARX model achieves the highest forecasting accuracy among the considered model types. Furthermore, for all models and specifications, the application of a variance-stabilizing transformation improves predictive performance, leading to systematically lower RMSE and MAE values. 

No single specification emerges as universally superior across all model classes. Instead, the optimal choice of estimator and modeled variable depends on the model type. For example, in the case of the NARX model, the pooled estimator yields the lowest forecast errors, whereas for ARX and LEAR the heterogeneous estimator performs best. Similarly, specifications based on deviations from the mean outperform direct models for ARX and NARX, but underperform for LEAR. These findings indicate that all considered model specifications contribute useful information to electricity price forecasting, with relative performance varying across model classes.

Overall, the models and specifications included in the forecasting pool exhibit substantial diversity, and no single model specification consistently dominates the results. Consequently, we find the pool  well suited for further analysis.

\begin{table}[]

\caption{Accuracy of ensemble forecasts across different model specifications}
\label{tab:rmse_mae}
\SetTblrInner{rowsep=0.5pt}
 \begin{tblr}{
  colspec={c|c|c|c|c|c},
  cell{3,10}{1} = {c=6}{c},
  cell{1}{1,2} = {r=2}{m},
  cell{4,6,8,11,13,15}{1} = {r=2}{m},
  cell{1}{3}= {c=4}{c}
}
Model& VST & Model specification & & & \\
\hline
 & &\hspace{0.5cm}Hetero \hspace{0.5cm}& \hspace{0.5cm}Pooled \hspace{0.5cm} & Hetero D-Mean & Pooled D-Mean \\

\hline
\hline 
RMSE & & & & & \\  
\hline
ARX  & None   & 28.99 & 29.2 & 28.96 & 29.17 \\
     & Asinh & 28.24 & 28.21 & \textbf{28.14} & 28.31 \\
\hline
NARX & None   & 26.15 & 27.33 & 26.26 & \textbf{25.55} \\
     & asinh& 25.68 & 25.71 & 26.28 & 25.72\\
\hline
LEAR & None   & 27.16 & 28.85 & 27.69 & 27.77 \\
     & Asinh &\textbf{26.06} & 27.96 & 26.61 & 26.8 \\
\hline
MAE & & & & & \\ 
\hline
ARX  & None    & 17.1 & 17.35 & 17.08 & 17.47 \\
ARX  & Asinh & 16.53 & 16.64 & \textbf{16.45} & 16.7\\
\hline
NARX & None    & 14.98 & 15.08 & 15.04 & \textbf{14.74} \\
NARX & Asinh & 14.75 & 15.07 & 14.95 & 14.76 \\
\hline
LEAR & None    & 15.72 & 17.72 & 16.26 & 16.55 \\
LEAR & Asinh & \textbf{14.99} & 16.76 & 15.43 & 15.71\\
\hline
\end{tblr}

\vspace{0.4cm}
\small{Note: The ensemble forecast is calculated as the average forecast over all considered estimation windows. The most accurate specification for each model type -- ARX, NARX and LEAR -- is highlighted in bold. }
\end{table}

\subsection{Profits}

The BESS income depends on the price spread within a day and the ability to select the charging and discharging hours. To disentangle these two features, we calculate first the profits of a BESS under the assumption of known future prices. The outcomes are called Oracle and are presented in the first column of Table \ref{tab:profit}. Next, we consider the income of a storage utility that uses imperfect forecasts from the pool in its decision process. Table  \ref{tab:profit} presents descriptive statistics of profits per 1 MWh of traded electricity. 

\begin{table}[h]
    \caption{Electricity prices and average daily profit per 1 MWh of traded electricity: descriptive statistics}
    \label{tab:profit}
    \centering
    \SetTblrInner{rowsep=0.5pt}
 \begin{tblr}{
  colspec={c|cc|c|cc|cc},
  cell{3,9}{1} = {c=8}{c},
  cell{1}{2} = {c=2}{c},
  cell{1}{4} = {c=5}{c},
  cell{1}{1} = {r=2}{m},
}
    Year & Prices & & Profits & & & &\\
    \hline
         & Mean & Std & Oracle &Max  & Min & Mean & Std\\
    \hline
    \hline
    BESS-a & & && & & &\\
    \hline
    2020 & 30.47 & 17.50 &1.42    & -0.48   & -4.24   & -1.46   & 0.81\\
    2021 & 96.85 & 73.68 &32.03   & 27.97   & 17.74   & 25.25   & 2.13\\
    2022 & 235.46& 142.80&105.84  & 96.53   & 77.04   & 91.88   & 4.18\\
    2023 & 95.18 & 47.58 &50.25   & 44.90   & 34.70   & 42.96   & 2.01\\
    2024 & 92.88 & 61.89 &92.57   & 80.43   & 62.58   & 77.15   & 3.25\\
    \hline
    BESS-b & & && & & &\\
    \hline
    2020 & 30.47 & 17.50 &-2.98   & -4.02   & -6.48  & -4.48   & 0.39\\
    2021 & 96.85 & 73.68 &23.23   & 20.84   & 16.32  & 19.27   & 1.08\\
    2022 & 235.46& 142.80&86.15   & 80.67   & 68.79  & 77.94   & 2.27\\
    2023 & 95.18 & 47.58 &38.13   & 35.49   & 28.94  & 33.88   & 1.10\\
    2024 & 92.88 & 61.89 &70.92   & 63.70   & 55.03  & 62.13   & 1.38\\
    \hline
    \end{tblr}
\end{table}

For both systems, BESS-a and BESS-b, {the results reveal a strong dependence of profitability on the analyzed year. In 2020, when electricity prices were relatively low and stable (with an average level of 30.47 and a standard deviation of 17.5), storage operators faced limited profit opportunities. Under perfect foresight, BESS-a achieved an average profit of 1.42 EUR/MWh, while BESS-b generated losses of -2.98 EUR/MWh. When imperfect price forecasts were used, average profits declined further to -1.46 EUR/MWh and -4.48 EUR/MWh, respectively.
In contrast, in 2022 -- when electricity prices peaked and exhibited substantial volatility (with the mean price increasing to 235.46 and the standard deviation exceeding 140) -- profit opportunities increased markedly. Under the Oracle benchmark, profits for the BESS-a system jumped to 91.88 EUR/MWh.}

Furthermore, when evaluating both types of systems, it's apparent that BESS-a generates greater profits compared to BESS-b. When the battery can be charged or discharged within one hour, it allows for achieving a lower buy- and higher sell price. Hence, the profits of Oracle and the average income of the pool decrease together with the battery power rating. 

Finally, one could observe significant discrepancies between the highest and the lowest profits in the pool. In year 2020, the most profitable forecasts brought income higher by {3.76} EUR/MWh than the worst ones. In the following years, this gap for BESS-a widened, reaching {19.49} EUR/MWh and {17.85} EUR/MWh in 2022 and 2024, respectively. In these two years, the additional earnings represented between {25-28}\% of the minimum income. The outcomes for BESS-b exhibit the same pattern, highlighting the critical role of accurate forecast selection in the BESS decision-making.

\subsection{Forecast evaluation and its relationship to profits}

As demonstrated earlier, the selection of predictions used in the decision-making process significantly influences the profitability of BESS. This choice can be based on traditional accuracy metrics, such as RMSE and MAE,  or other measures associated with the behavior of forecast errors (Cov-e) or the association between the predictions and true values (Corr-f). In this section, the relationship between measures of forecast quality and battery profits is evaluated. It can be noticed that using individual predictions, we can calculate both the income level and a measure value. Hence, there are 192 (for a big pool) or 64 (for model-based pools) pairs of values that describe the outcomes. Using this information, we calculate the Spearman correlation, which is robust to outliers and any monotonic transformation of data. 

In Table \ref{tab:Corr},  correlation coefficients between forecast quality measures and the average level of profits calculated for the whole evaluation period are presented. The first block of outcomes shows the results for BESS-a, whereas the second block represents BESS-b. The results are displayed for three sub-pools and for a big pool separately. The findings demonstrates that all correlations, apart from Corr-f, are negative. This implies that an increase in, for instance, Cov-e, is associated with a reduction in profits. Unlike other metrics, Corr-f is interpreted positively, meaning that a higher value of this measure corresponds to a greater income. Because the sign of the correlation coefficient varies across measures, we refer to its absolute value  when comparing the strength of correlations in the following sections of the article.

\begin{table}[h]
    \caption{The average Spearman correlation coefficient between profits and forecast quality measures; rolling window of 365 days }
    \label{tab:Corr}
    \centering
        \SetTblrInner{rowsep=0.5pt}
 \begin{tblr}{
  colspec={c|ccc|c},
  cell{2,9}{1} = {c=5}{c}
}
    Measure & ARX & NARX  & LEAR & All\\
    \hline
    \hline
    BESS-a& & & & \\
    \hline
 RMSE      & -0.39  & -0.29  & -0.08  & -0.20   \\
 MAE       & -0.40  & -0.39  & -0.02  & -0.20  \\
 \hline
Cov-e      & -0.72  & -0.85  & -0.71  & -0.77  \\
Corr-f     &  0.90  &  0.85  & 0.87   & 0.86  \\
\hline
MHD        & -0.83  & -0.82  & 0.78   & -0.79 \\
MPD        & -0.97  & -0.96  & -0.96  & -0.97 \\
\hline
 BESS-b & & & & \\
\hline
RMSE      & -0.46  & -0.38  & -0.18  & -0.34   \\
 MAE      & -0.47  & -0.44  & -0.13  & -0.35  \\
 \hline
Cov-e     & -0.70  & -0.76  & -0.66  & -0.69  \\
Corr-f    & 0.92   &  0.86  &  0.89  & 0.90 \\
\hline
MHD       & -0.86  & -0.79  & -0.82  & -0.83  \\
MPD       & -0.93  & -0.84  & -0.90  & -0.89\\
\hline
    \end{tblr}
\end{table}

Let us first analyze the results for BESS-a system and the big pool of forecasts. {The analysis indicates} that widely used accuracy metrics, such as RMSE and MAE, show weak correlation with profits, with the absolute correlation coefficient being {close}, in absolute terms, to {0.20}. At the same time, correlation with dispersion and association measures {oscillates around 0.80}. This highlights a substantial difference between these evaluation methods. The performance of Corr-f measure, which shows how well the forecasts reflect the within-day fluctuations of electricity prices, is worth emphasizing. The value of the correlation between Corr-f and average profits reaches {0.855}. 

{An analysis of smaller, model-based forecast pools shows that, while the overall correlation pattern remains similar, the strength of the correlations varies across pools.} The highest correlation is observed for the ARX and NARX models, which, notably, generate the most diverse set of predictions. For NARX, the correlation coefficient ranges from {0.29 to 0.96}. In contrast, the weakest association is found with the LEAR models, where the correlation between RMSE or MAE and profits is nearly zero.

The results for BESS-b closely resemble  those of BESS-a. Traditional accuracy metrics, such as RMSE and MAE, show only a weak correlation with profits; however, this relationship is slightly stronger compared to BESS-a.  The overall outcomes suggest that the decision-making process in BESS-b is more complex than in BESS-a case. As a result, the financial performance of BESS-b  is more dependent on accurately capturing the shape of the daily price curve -- an aspect measured by Corr-f.

Finally, we examine the correlation between profits and two specialized measures designed to assess the accuracy of selecting the minimum and maximum price hours within a day: MHD and MPD. 
The MHD metric penalizes deviations from the optimal charging and discharging hours. While it ranks among the better-performing indicators, its correlation with profits is weaker than that of Corr-f and comparable to Cov-e. In contrast, MPD demonstrates a significantly stronger relationship with profits, showing an average correlation of {0.96} for BESS-a and {0.89} for BESS-b. Unlike MHD, which measures time-based discrepancies, MPD assesses the difference of electricity prices between the selected and optimal hours. Given that profits are directly tied to price differences -- and that the daily price curve often exhibits a "duck" shape -- MPD proves to be a more effective performance metric than MHD.

\begin{figure}[p!]
    \centering
    \includegraphics[width=0.85\textwidth]{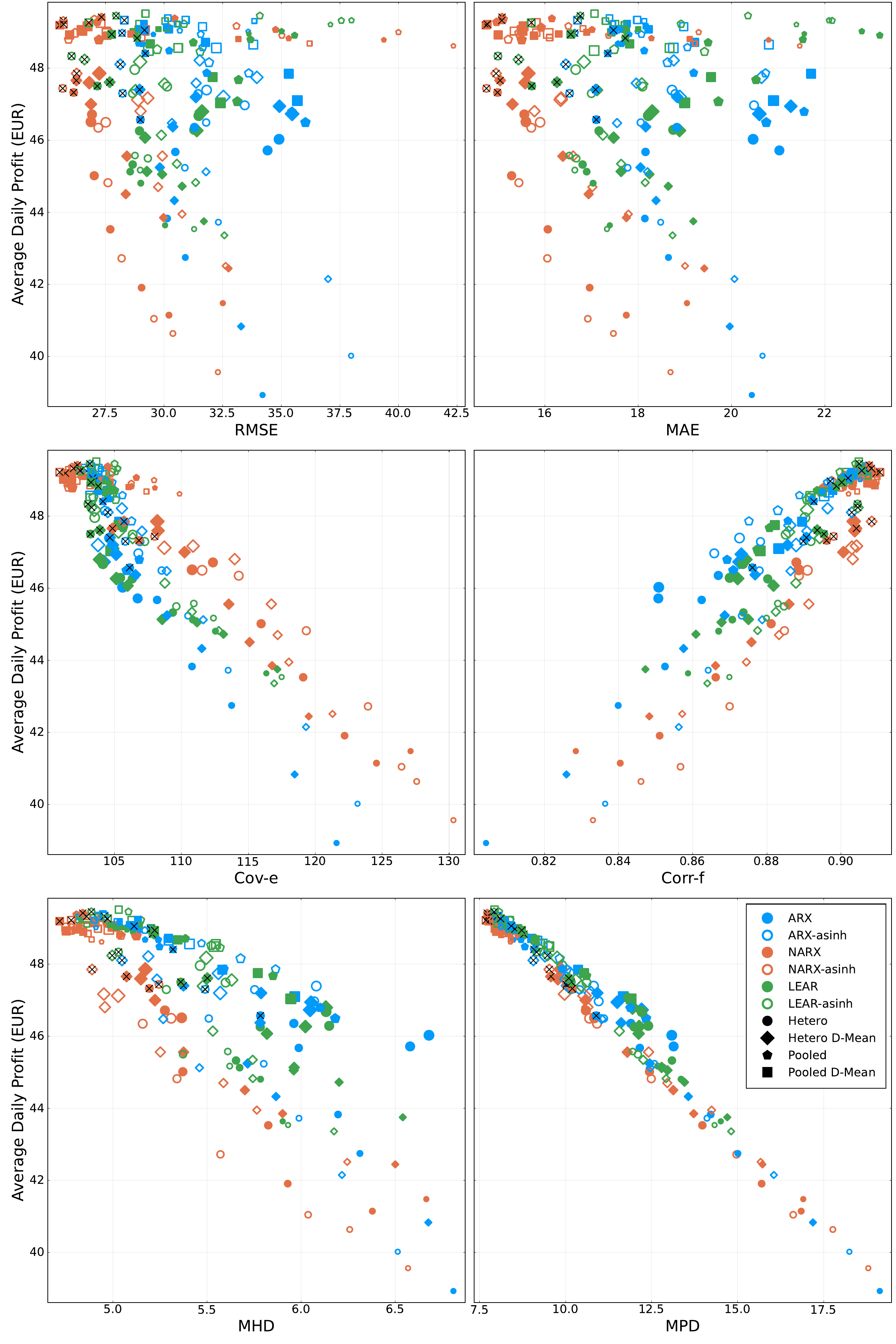}

    \caption{Scatter plots of average daily profits for the BESS-a energy storage calculated for the entire testing set with respect to different error measures.}
    \label{fig:ScatterProfits}
\end{figure}

A more detailed illustration of the relationship between forecast quality measures and profitability is presented in Fig.~\ref{fig:ScatterProfits}. The figure depicts scatter plots of average daily income per 1 MWh of trade ($\pi$) against various forecast evaluation metrics for the BESS-a energy storage system. In these plots, colors denote different models (blue - ARX, red - NARX, green - LEAR), shapes indicate the estimator type and model specification, and marker size reflects the length of the calibration window. Finally, a cross indicates the average of forecasts across different estimation windows.

The results confirm earlier findings of a weak correlation between RMSE/MAE and profits (top row, Fig.~\ref{fig:ScatterProfits}). The scatter plots form diffuse clouds with no clear monotonic relationship, with one noticeable feature: a cluster of red points corresponding to NARX models. These results indicate that NARX provides the most accurate forecasts; however, its associated income is comparable to -- or in some cases lower than -- that of alternative models.
In contrast, the Cov-e and Corr-f measures (middle row, Fig.~\ref{fig:ScatterProfits}) demonstrate a substantially stronger and nearly linear relationship with income. For the Corr-f metric in particular, outcomes cluster distinctly by model, with LEAR consistently yielding the highest profits.

Finally, when comparing model specifications, the pooled estimator consistently outperforms the heterogeneous estimator, delivering forecasts that combine high predictive accuracy with strong profitability. {This finding suggests that the heterogeneous estimation approach, which dominates much of the existing literature, should not necessarily be regarded as the superior forecasting strategy.
Several factors may explain the strong performance of the pooled estimator. First, since model parameters for all hours are calibrated jointly, the resulting estimates tend to be more stable and less sensitive to outliers. Second, unexpected changes in market conditions -- such as shocks driven by intermittent RES generation -- are likely to affect predictions of prices across all hours of the day in the same way. In such cases, pooling reduces the variability of forecast errors across hours, which in turn leads to more consistent and accurate trading decisions.}

\subsubsection{Time evolution of correlation}

To account for changes in correlation over time, a rolling window method is applied. The process begins by selecting the initial 365 days in the evaluation window. Within this set, the average profit is computed along with six different forecast quality metrics and their corresponding correlation coefficients. Next, the window is shifted by one day, and the whole procedure is repeated. 
The time evolution of the resulting correlation coefficients is illustrated in Fig.~\ref{fig:Corr:BESS} for BESS-a (upper panel)  and BESS-b (lower panel), respectively. {The shaded regions represent 95\% confidence intervals around the coefficients and illustrate the uncertainty associated with the estimation process.} For visualization purposes, the coefficients of all metrics except Corr-f are multiplied by –1.

\begin{figure}[tb]
    \centering
    \begin{subfigure}[a]{\textwidth}
        \captionsetup[sup]{}
        \caption*{BESS-a}
        \centering
        \includegraphics[width=\textwidth]{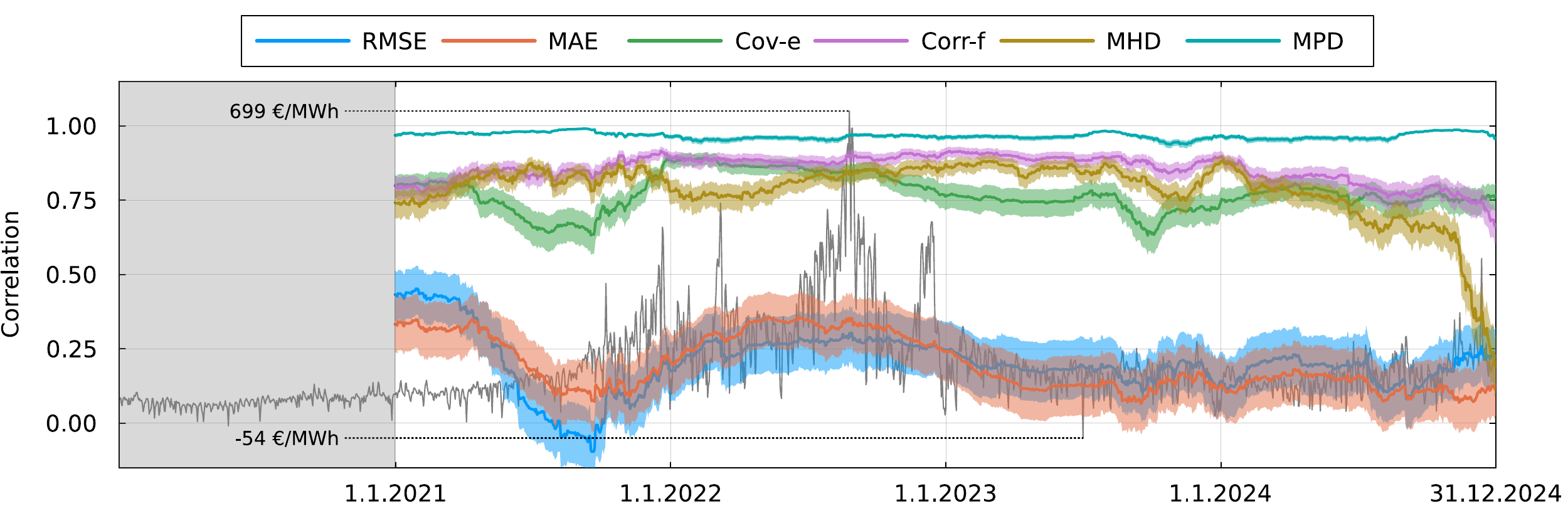}
    \end{subfigure}
    \begin{subfigure}[b]{\textwidth}
        \captionsetup[sup]{}
        \caption*{BESS-b}
        \centering
        \includegraphics[width=\textwidth]{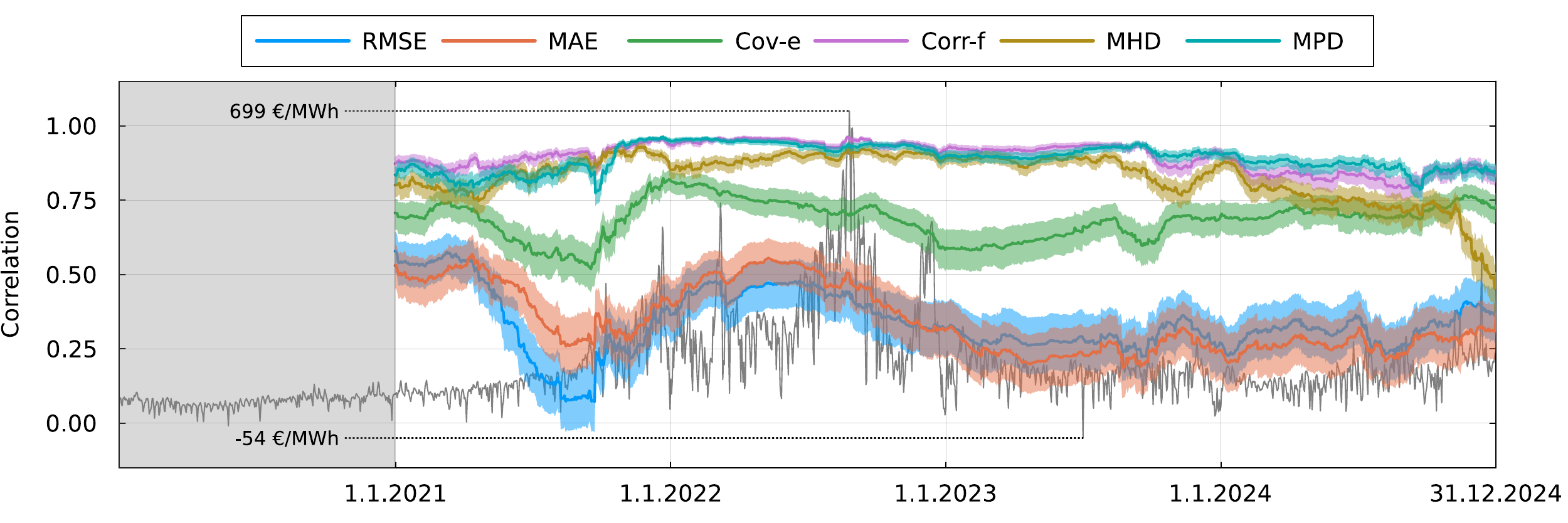}
    \end{subfigure}
    \caption{Spearman correlation between forecast quality measures and profits for the BESS-a (top panel) and BESS-b (bottom panel) energy storage. The grey line marks the average daily price of electricity.}
    \label{fig:Corr:BESS}
\end{figure}

The results reveal a clear downward trend in the correlation between RMSE or MAE and profits. Not only is the relationship between these traditional accuracy measures and income the weakest overall, but it also deteriorates over time. At the start of the evaluation period, which includes data from 2020, the  correlation between MAE and profit oscillated around {\nobreakdash-0.35}. By 2021, it drops for BESS-a to around -0.06, and in the final two years, it fluctuates between {-0.10}. RMSE exhibits a similar trend, with a noticeable decline in correlation over the years. This progression supports earlier conclusions that RMSE and MAE are poor indicators of profitability.

When examining the performance of  a measure that captures error dispersion, several similar patterns emerge. The correlation between profits and Cov-e declined in absolute terms as electricity prices began to rise in 2021. Subsequently, these correlations recovered, even slightly surpassing their initial levels. For the BESS-a specification, the strength of correlation remained relatively strong, whereas for the BESS-b specification, it fluctuates and declines slightly over time.

In contrast to the accuracy and dispersion measures, the performance of the association measure remains stable over most of the evaluation period, oscillating around the average value reported in Table~\ref{tab:Corr}. Only from 2023H2 the correlation between Corr-f and profits starts to fall below {0.80}. On the plot it is visible as a drop in rolling correlation in 2024H2. {However, this weakening is only slight for BESS-a, where the coefficient declines to 0.69 by the end of the sample, while for BESS-b it remains consistently close or above 0.80.}

Finally, let us consider the results for MHD and MPD, which confirm earlier findings of the inferior performance of the hour-based metric. The correlation between MHD and profits consistently remains lower than that of MPD. Notably, there is a sharp decline in MHD performance toward the end of the sample period -- dropping from {0.85} to {0.20} within a single year for BESS-a. This decline is associated with the changing shape of daily price curves. As more and more solar energy enters the system, the midday reduction of electricity prices becomes more pronounced, resulting in not just two peaks (morning and evening), but also two low-price periods (night and midday). Consequently, the lowest and highest prices are not clustered closely to each other. For example, if the actual minimum price occurs at night and the forecasted minimum is during midday, despite similar price levels, the time-based error is heavily penalized. Thus, evaluating performance based on price differences rather than hours proves to be a more robust method. The absolute value of the correlation between MPD and profits remains above {0.77} throughout the evaluation period for both BESS-a and BESS-b, and for BESS-a, it consistently exceeds {0.90}.

 A comparison of the upper and the lower panel in Fig.~\ref{fig:Corr:BESS} reveals that both battery systems exhibit a similar overall pattern in correlation coefficients. However, some subtle differences can be observed. For BESS-b, the correlation coefficients related to dispersion and association measures tend to be more stable compared to BESS-a. Additionally, the performance of  Corr-f improves in BESS-b, approaching the level of MPD, unlike in the first system. Lastly, as indicated in Table \ref{tab:Corr}, while traditional accuracy measures still show the weakest correlation with profits, their relationship is slightly stronger in BESS-b than in BESS-a.

{
\section{Discussion}\label{sec:Discussion}

\subsection{Point forecasts and probabilistic forecasting}
While modern electricity markets increasingly adopt probabilistic and quantile-based forecasts, this study deliberately focuses on point forecasts. In many practical applications, particularly in day-ahead market participation and deterministic optimization of battery energy storage systems, point forecasts remain the primary input for operational decision-making. Focusing on point predictions allows us to isolate the relationship between forecast quality and economic value without introducing additional layers related to risk preferences, uncertainty modeling, or stochastic decision rules.

Importantly, the objective of this paper is not to compare point and probabilistic forecasting paradigms, but to examine how different properties of point forecasts translate into economic performance under a fixed trading strategy. Extending the proposed evaluation framework to probabilistic forecasts, by incorporating distribution-based metrics or stochastic optimization of BESS operation, constitutes a natural and relevant direction for future research, but represents a distinct research question beyond the scope of the present study. Another open problem is expanding the analysis to other machine learning models, possibly including gradient boosted decision trees, deep neural networks and transformer-based architectures.

\subsection{Interpretation of MHD and MPD measures}
{
We consider two measures that explicitly assess the identification of daily minimum and maximum price hours: MHD and MPD. The former quantifies the temporal deviation between predicted and realized peak or trough hours, while the latter evaluates the accuracy of hour selection based on differences in the corresponding price levels. The results indicate that hour-based assessment can be misleading, particularly when the rising share of solar generation changes the daily pattern of electricity prices.

Although MPD exhibits a strong correlation with BESS profits, it relies exclusively on two price points and therefore provides only a coarse proxy for arbitrage opportunities, especially for storage systems operating over multi-hour charging and discharging blocks. {This limitation motivates the emphasis placed in this study on dispersion- and association-based measures such as Cov-e and Corr-f, which are based on the full intraday price profile and are more informative for block-based storage operation.}
}

\subsection{Assumptions and limitations}

The results of this study should be interpreted in light of several simplifying assumptions. First, because the primary objective of this research is not to replicate the detailed operational behavior of the energy storage system but rather to evaluate the relationship between forecast quality measures and arbitrage value, the decision problem associated with the BESS is intentionally simplified. Specifically, we assume that the battery completes one full charge–discharge cycle per day and does not transfer energy between consecutive days. Furthermore, no additional constraints on the minimum or maximum state of charge are imposed, although such limits are typically required in real-world applications to extend battery lifetime. Finally, given the relatively short evaluation period considered in this study, battery degradation is not modeled explicitly within the decision problem; instead, it is indirectly reflected through reduced BESS efficiency. Allowing for multiple charge–discharge cycles within a single day and introducing additional operational constraints would require the use of more complex optimization methods, such as linear programming, mixed-integer linear programming, or nonlinear programming \citep[see][for a discussion]{mer:etal:23}. Implementing such approaches would substantially increase computational time, particularly when evaluating a pool of 192 forecasts, but would add little to final conclusions.

Second, the analysis is restricted to day-ahead market trading and does not consider intraday adjustments, balancing markets, or ancillary services. While day-ahead arbitrage remains a core revenue stream for many storage operators, additional market layers could alter both optimal trading strategies and the relative importance of different forecast properties.

Finally, the empirical analysis is conducted for the German electricity market, which operates as a single bidding zone with relatively high liquidity. In markets with stronger congestion effects, zonal fragmentation, or higher price volatility, the relationship between forecast quality measures and economic value may differ. These aspects provide natural directions for extending the proposed framework to other market designs and operating conditions.
}

\section{Conclusions}
\label{sec:conclusion}

{This paper examines the relationship between statistical forecast quality and economic value in electricity markets, with a particular focus on BESS arbitrage.} We consider two hypothetical BESS configurations that have the same energy capacity but differ in power ratings. This difference significantly affects financial outcomes, as a battery with a lower power rating requires more time to fully charge or discharge, potentially forcing it to operate during less favorable price periods.

In this article, we construct a comprehensive pool of 192 forecasts based on three widely used model types: ARX, LEAR, and NARX, which together encompass a broad family of linear and nonlinear approaches. Additionally, we consider eight different model specifications that vary according to the choice of the endogenous variable, the use of variance-stabilizing transformations, and the type of estimator (heterogeneous vs. pooled). Model parameters are estimated using rolling windows of seven different lengths. The pool is further enriched with ensemble forecasts, which average predictions across different window sizes.

{
The empirical findings demonstrate that forecast properties beyond point accuracy, as measured by RMSE and MAE, are crucial for economic performance. In particular, measures capturing intraday error dispersion (Cov-e) and the association between predicted and realized prices (Corr-f) exhibit a strong relationship with BESS profitability. These metrics reflect a forecast’s ability to reproduce the intraday price profile and to correctly identify profitable charging and discharging opportunities. In contrast, relying solely on RMSE or MAE may result in the selection of forecasts that are statistically accurate yet economically suboptimal.

The major results of this research can be summarized as follows:}
\begin{itemize}
    \item The choice of forecasting model has a substantial influence on BESS profitability. Selecting the best-performing model can increase profits by {16–58}\% during 2021–2024 compared to the poorest-performing model.
    \item {Among the analyzed model classes, NARX delivers the highest statistical accuracy, while LEAR-based forecasts generate the highest profits. With respect to model specification, the pooled estimator emerges as an attractive alternative to the commonly used heterogeneous approach.} 
    \item The weak correlations between MAE and RMSE and profits, spanning {–0.45 to 0.10 for BESS-a and -0.58 to -0.07 for BESS-b}, indicate that these metrics are insufficient for reliably differentiating forecast performance.
    \item {The correlation between the dispersion measure, Cov-e, and profits ranges from -0.89 to  -0.63, for BESS-a. Similarly,  Corr-f, shows a strong correlation with profits, oscillating between 0.66 and 0.92, substantially outperforming traditional accuracy metrics.}
    \item Among the two measures designed to assess the accuracy of identifying hours of minimum and maximum prices, MPD exhibits a stronger correlation with profits than MHD.
\end{itemize}

{
We believe that these findings contribute to the ongoing discussion on forecast evaluation by highlighting the importance of explicitly linking forecast quality to economic value, particularly in the context of electricity markets. Several area for future research emerge from this study. First, the proposed metrics -- particularly MPD -- can enrich forecast evaluation frameworks, especially when assessing newly developed forecasting methods. Moreover, they could be used for forecast selection or for assigning weights in forecast ensembles. In the existing literature, such weights are typically derived from RMSE or MAE, or estimated via regression-based approaches. In this regard, the Corr-f measure offers an attractive alternative, as it can take both positive and negative values and is naturally bounded within the interval $(-1,1)$.

Finally, some of the proposed measures, such as Cov-e and Corr-f, are smooth and differentiable, which makes them suitable candidates for inclusion in loss functions to enhance the estimation process. It should be noted, however, that such an approach would require the joint estimation of parameters across all 24 hourly models. Moreover, since Corr-f is invariant to linear transformations of forecasts, it must be combined with accuracy-based measures, such as the mean squared error, to ensure model identification. The potential benefits of augmenting standard loss functions, such as the sum of squared errors, with Cov-e- or Corr-f-based components are not immediate and require further investigation.
}

\section*{CRediT authorship contribution statement}

\textbf{Katarzyna Maciejowska:} Conceptualization, Formal analysis, Methodology, Software, Investigation, Funding acquisition, Supervision, Validation, Writing – original draft
\textbf{Arkadiusz Lipiecki:} Formal analysis, Methodology, Software, Investigation, Visualization, Writing – original draft 
\textbf{Bartosz Uniejewski:} Formal analysis, Methodology, Software, Investigation, Funding acquisition, Supervision, Validation, Writing – original draft

\section*{Declaration of competing interests}
The authors declare no competing interests.

\section*{Acknowledgements}
The study was partially supported by the National Science Center (NCN, Poland) through grant no.\ 2018/30/A/HS4/00444 (to AL), grant no.\ 2023/49/N/HS4/02741 (to BU) and grant no.\ 2019/34/E/HS4/00060 (to KM).

\bibliographystyle{elsarticle-num} 
\bibliography{ref}

@article{tsc:etal:2022,
    title = {Electricity price forecasting on the day-ahead market using machine learning},
    journal = {Applied Energy},
    volume = {313},
    pages = {118752},
    year = {2022},
    issn = {0306-2619},
    author = {Léonard Tschora and Erwan Pierre and Marc Plantevit and Céline Robardet}
}

@article{jed:etal:2022,
  author={Jędrzejewski, Arkadiusz and Lago, Jesus and Marcjasz, Grzegorz and Weron, Rafał},
  journal={IEEE Power and Energy Magazine}, 
  title={Electricity Price Forecasting: The Dawn of Machine Learning}, 
  year={2022},
  volume={20},
  number={3},
  pages={24-31},
}

@article{yam:etal:22,
title = {Multi-service based economic valuation of grid-connected battery energy storage systems},
journal = {Journal of Energy Storage},
volume = {52},
pages = {104657},
year = {2022},
author = {Sumanth Yamujala and Anjali Jain and Rohit Bhakar and Jyotirmay Mathur},
}

@article{par:etal:25,
    title = {On the participation of energy storage systems in reserve markets using Decision Focused Learning},
    journal = {Sustainable Energy, Grids and Networks},
    volume = {42},
    pages = {101677},
    year = {2025},
    author = {Ángel Paredes and Jean-François Toubeau and José A. Aguado and François Vallée},
}

@article{alk:etal:26,
    title = {Decision-Focused Learning Enhanced by Automated Feature Engineering for Energy Storage Optimisation},
    journal = {Expert Systems with Applications},
    volume = {302},
    pages = {130554},
    year = {2026},
    author = {Nasser Alkhulaifi and Ismail {Gokay Dogan} and Timothy R. Cargan and Alexander L.     Bowler and Direnc Pekaslan and Nicholas J. Watson and Isaac Triguero}
}

@book{sch:sta:23,
    author = {Oliver Schmidt and Iain Staffell},
    title = {Monetizing energy storage},
    publisher = {Oxford University Press},
    year = {2023}
}

@article{zar:can:bha:10,
  author =        {Zareipour, Hamidreza and Canizares, Claudio A. and
                   Bhattacharya, Kankar},
  journal =       {IEEE Transactions on Power Systems},
  number =        {1},
  pages =         {254-262},
  title =         {Economic Impact of Electricity Market Price
                   Forecasting Errors: A Demand-Side Analysis},
  volume =        {25},
  year =          {2010},
}

@article{lin:zhu:wid:24,
  author =        {O. Lindberg and R. Zhu and J. Widén},
  journal =       {Renewable Energy},
  pages =         {121617},
  title =         {Quantifying the value of probabilistic forecasts when
                   trading renewable hybrid power parks in day-ahead
                   markets: A Nordic case study},
  volume =        {237},
  year =          {2024},
}

@article{car:kar:19,
  author =        {Carriere, Thomas and Kariniotakis, George},
  journal =       {IEEE Transactions on Smart Grid},
  number =        {6},
  pages =         {6933-6944},
  title =         {An Integrated Approach for Value-Oriented Energy
                   Forecasting and Data-Driven Decision-Making
                   Application to Renewable Energy Trading},
  volume =        {10},
  year =          {2019},
}

@article{keb:ara:rah:2011,
  author =        {Kebriaei, Hamed and Araabi, Babak N. and
                   Rahimi-Kian, Ashkan},
  journal =       {IEEE Transactions on Power Systems},
  number =        {4},
  pages =         {1817-1825},
  title =         {Short-Term Load Forecasting With a New Nonsymmetric
                   Penalty Function},
  volume =        {26},
  year =          {2011},
}

@article{li:chi:2018,
  author =        {Li, Guo and Chiang, Hsiao-Dong},
  journal =       {IEEE Transactions on Smart Grid},
  number =        {4},
  pages =         {2508-2517},
  title =         {Toward Cost-Oriented Forecasting of Wind Power
                   Generation},
  volume =        {9},
  year =          {2018},
}

@article{zha:wan:hug:2022,
  author =        {Jialun Zhang and Yi Wang and Gabriela Hug},
  journal =       {Electric Power Systems Research},
  pages =         {107723},
  title =         {Cost-oriented load forecasting},
  volume =        {205},
  year =          {2022},
}

@article{ser:wer:2024,
title = {Loss functions in regression models: Impact on profits and risk in day-ahead electricity trading},
journal = {Energy Economics},
volume = {148},
pages = {108596},
year = {2025},
author = {Tomasz Serafin and Rafał Weron},
}

@article{Zha:23,
  author =        {Yufan Zhang and Mengshuo Jia and Honglin Wen and
                   Yuexin Bian and Yuanyuan Shi},
  journal =       {IEEE Transactions on Smart Grid},
  title =         {Toward Value-Oriented Renewable Energy Forecasting:
                   An Iterative Learning Approach},
  year =          {2023},
}

@article{mand:etal:24,
  author =        {Mandi, Jayanta and Kotary, James and BErden, Senne and
                   Mulamba, Maxime and Bucarey, Victor and Guns, Tias and
                   Fiorett, Ferdinando},
  journal =       {Journal of Artificial Intelligence Research},
  pages =         {1623-1701},
  title =         {Decision-focused learning: foundations, state of the
                   art, benchmark and future opportunities},
  volume =        {81},
  year =          {2024},
}

@article{mur:1993,
  author =        {Allan H. Murphy},
  journal =       {Weather and Forecasting},
  number =        {2},
  pages =         {281 - 293},
  title =         {What Is a Good Forecast? An Essay on the Nature of
                   Goodness in Weather Forecasting},
  volume =        {8},
  year =          {1993},
}

@article{zie:wer:18,
  author =        {F. Ziel and R. Weron},
  journal =       {Energy Economics},
  pages =         {396-420},
  title =         {Day-ahead electricity price forecasting with
                   high-dimensional structures: {U}nivariate vs.
                   multivariate modeling frameworks},
  volume =        {70},
  year =          {2018},
}

@article{lag:mar:des:wer:21,
  author =        {Lago, J. and Marcjasz, G. and De Schutter, B. and
                   Weron, R.},
  journal =       {Applied Energy},
  pages =         {116983},
  title =         {Forecasting day-ahead electricity prices: A review of
                   state-of-the-art algorithms, best practices and an
                   open-access benchmark},
  volume =        {293},
  year =          {2021},
}

@article{mac:uni:wer:23,
  author =        {K. Maciejowska and B.Uniejewski and R. Weron},
  journal =       {Oxford research encyclopedia of economics and
                   finance},
  title =         {Forecasting Electricity Prices},
  year =          {2023},
}

@article{mar:etal:23,
  author =        {Grzegorz Marcjasz and Michał Narajewski and
                   Rafał Weron and Florian Ziel},
  journal =       {Energy Economics},
  pages =         {106843},
  title =         {Distributional neural networks for electricity price
                   forecasting},
  volume =        {125},
  year =          {2023},
}

@article{mar:uni:wer:19:narx,
  author =        {G. Marcjasz and B. Uniejewski and R. Weron},
  journal =       {International Journal of Forecasting},
  pages =         {1520-1532},
  title =         {On the importance of the long-term seasonal component
                   in day-ahead electricity price forecasting with
                   {NARX} neural networks},
  volume =        {35},
  year =          {2019},
}

@article{hub:mar:wer:19,
  author =        {K. Hubicka and G. Marcjasz and R. Weron},
  journal =       {IEEE Transactions on Sustainable Energy},
  number =        {1},
  pages =         {321-323},
  title =         {A note on averaging day-ahead electricity price
                   forecasts across calibration windows},
  volume =        {10},
  year =          {2019},
}

@article{mar:uni:wer:20,
  author =        {G. Marcjasz and B. Uniejewski and R. Weron},
  journal =       {International Journal of Forecasting},
  pages =         {466-479},
  title =         {Probabilistic electricity price forecasting with
                   {NARX} networks: Combine point or probabilistic
                   forecasts?},
  volume =        {36},
  year =          {2020},
}

@article{hag:men:94,
  author =        {Hagan, M.T. and Menhaj, M.B.},
  journal =       {IEEE Transactions on Neural Networks},
  number =        {6},
  pages =         {989-993},
  title =         {Training feedforward networks with the Marquardt
                   algorithm},
  volume =        {5},
  year =          {1994},
}

@article{tib:96,
  author =        {R. Tibshirani},
  journal =       {Journal of the Royal Statistical Society B},
  pages =         {267-288},
  title =         {Regression shrinkage and selection via the lasso},
  volume =        {58},
  year =          {1996},
}

@article{uni:24:ORD,
  author =        {B. Uniejewski},
  journal =       {Operations Research and Decisions},
  title =         {Regularization for electricity price forecasting},
  volume =        {34},
  year =          {2024},
}

@article{mar:ser:wer:18,
  author =        {G. Marcjasz and T. Serafin and R. Weron},
  journal =       {Energies},
  pages =         {2364},
  title =         {Selection of Calibration Windows for Day-Ahead
                   Electricity Price Forecasting},
  volume =        {11},
  year =          {2018},
}

@article{ser:uni:wer:19,
  author =        {Serafin, T. and Uniejewski, B. and Weron, R.},
  journal =       {Energies},
  number =        {13},
  pages =         {256},
  title =         {Averaging predictive distributions across calibration
                   windows for day-ahead electricity price forecasting},
  volume =        {12},
  year =          {2019},
}

@article{gia:gro:12,
  author =        {A. Gianfreda and L. Grossi},
  journal =       {Energy Economics},
  number =        {6},
  pages =         {2228-2239},
  title =         {Forecasting {I}talian electricity zonal prices with
                   exogenous variables},
  volume =        {34},
  year =          {2012},
}

@article{cia:mun:zar:22,
  author =        {Ciarreta, Aitor and Muniain, Peru and
                   Zarraga, Ainhoa},
  journal =       {Electric Power Systems Research},
  pages =         {108144},
  publisher =     {Elsevier},
  title =         {Do jumps and cojumps matter for electricity price
                   forecasting? Evidence from the German-Austrian
                   day-ahead market},
  volume =        {212},
  year =          {2022},
}

@article{jan:tru:wer:wol:13,
  author =        {J. Janczura and S. Tr\"uck and R. Weron and R. Wolff},
  journal =       {Energy Economics},
  pages =         {96-110},
  title =         {Identifying spikes and seasonal components in
                   electricity spot price data: A guide to robust
                   modeling},
  volume =        {38},
  year =          {2013},
}

@article{uni:wer:zie:18,
  author =        {B. Uniejewski and R. Weron and F. Ziel},
  journal =       {IEEE Transactions on Power Systems},
  pages =         {2219-2229},
  title =         {Variance stabilizing transformations for electricity spot price forecasting},
  volume =        {33},
  year =          {2018},
}

@book{arm:01,
  author =        {J.S. Armstrong},
  publisher =     {Springer},
  title =         {Principles of Forecasting: A handbook for researchers and practitioners},
  year =          {2001},
}

@incollection{tim:06,
  author =        {A. G. Timmermann},
  booktitle =     {Handbook of economic forecasting},
  editor =        {G. Elliott and C. W. Granger and A. Timmermann},
  pages =         {135-196},
  publisher =     {Elsevier},
  title =         {Forecast combinations},
  year =          {2006},
}

@article{ati:2020,
  author =        {Amir F. Atiya},
  journal =       {International Journal of Forecasting},
  number =        {1},
  pages =         {197-200},
  title =         {Why does forecast combination work so well?},
  volume =        {36},
  year =          {2020},
}

@article{mac:now:wer:16,
  author =        {K. Maciejowska and J. Nowotarski and R. Weron},
  journal =       {International Journal of Forecasting},
  number =        {3},
  pages =         {957-965},
  title =         {Probabilistic forecasting of electricity spot prices using {F}actor {Q}uantile {R}egression {A}veraging},
  volume =        {32},
  year =          {2016},
}

@article{lip:etal:2024,
  author =        {Arkadiusz Lipiecki and Bartosz Uniejewski and
                   Rafał Weron},
  journal =       {Energy Economics},
  pages =         {107934},
  title =         {Postprocessing of point predictions for probabilistic
                   forecasting of day-ahead electricity prices: The
                   benefits of using isotonic distributional regression},
  volume =        {139},
  year =          {2024},
}

@article{ama:etal:2023,
  author =        {Yvenn Amara-Ouali and Matteo Fasiolo and Yannig Goude and
                   Hui Yan},
  journal =       {International Journal of Forecasting},
  number =        {3},
  pages =         {1272-1286},
  title =         {Daily peak electrical load forecasting with a
                   multi-resolution approach},
  volume =        {39},
  year =          {2023},
}

@article{sha:26,
    title = {Adaptive optimal operation of grid-connected battery systems under varying electricity market volatility},
    journal = {Energy Conversion and Management},
    volume = {348},
    pages = {120596},
    year = {2026},
    author = {Masoume Shabani}
}

@article{mer:etal:23,
title = {The value of electricity storage arbitrage on day-ahead markets across {E}urope},
journal = {Energy Economics},
volume = {123},
pages = {106721},
year = {2023},
author = {Thomas Mercier and Mathieu Olivier and Emmanuel {De Jaeger}},
}

\end{document}